# Time-Reversal Symmetry Protected Transport at Correlated Oxide Interfaces

Mengke Ha[1*], Qing Xiao[1*], Zhiyuan Qin[1,2*], Dawei Qiu[1], Longbing Shang[1], Xinyi Liu[4], Pu Yan[5], Changjian Ma[1], Danqing Liu[1], Chengyuan Huang[1], Zhenlan Chen[1,2], Haoyuan Wang[1], Chang-Kui Duan[1,2], Zhaoliang Liao[6], Wei-Tao Liu[4], Yang Gao[7], Kecheng Cao[5], Jiangfeng Du[1,2,3†], and Guanglei Cheng[1,2†]

[1]CAS Key Laboratory of Microscale Magnetic Resonance and School of Physical Sciences, University of Science and Technology of China, Hefei 230026, China
[2]Hefei National Laboratory, University of Science and Technology of China，Hefei 230026, China
[3]Institute of Quantum Sensing and School of Physics, Zhejiang University, Hangzhou 310058, China
[4]Physics Department, State Key Laboratory of Surface Physics, Key Laboratory of Micro and Nano Photonic Structures, Fudan University, Shanghai 200433, China
[5]School of Physical Science and Technology & Shanghai Key Laboratory of High-resolution Electron Microscopy, ShanghaiTech University, Shanghai 200093, China
[6]National Synchrotron Radiation Laboratory, University of Science and Technology of China, Hefei 230026, China
[7]Department of Physics, University of Science and Technology of China, Hefei 230026, China

**Abstract:** Time-reversal symmetry (TRS) protection is core to topological physics, yet its role in correlated oxides-typically non-topological-remains underexplored. This limit hampers the potential in engineering exotic quantum states by fusing TRS protection and the rich emergent phenomena in the oxide platform. Here, we report evidence of a TRS-protected subband at oxygen vacancy-free $LaAlO_3/SrTiO_3$ interfaces. This subband causes a low-field quantum oscillation with anomalous characters: exceptionally light electron mass, aperiodicity, and susceptibility to magnetic fields. All findings align with a Rashba model in which TRS-protected transport occurs along quasi-1D ferroelastic domain walls, which possess a Dirac band topology and a giant Rashba spin-orbit coupling, two orders stronger than the 2D interface. Our results deepen the understanding of $SrTiO_3$-based electron systems, unveiling an appealing new platform for quantum engineering.

## INTRODUCTION

Time-reversal symmetry (TRS) is fundamental to many novel physical phenomena, including topological surface states [1,2], quantum magnetism [3], and unconventional superconductivity [4]. It protects transport from local perturbations by imposing spin-momentum locking in quantum materials with strong spin-orbit coupling (SOC) [5,6]. It is, therefore, a crucial ingredient in engineering quantum states of matter, e.g., the Majorana zero mode [7].

*glcheng@ustc.edu.cn

Meanwhile, correlated oxide interfaces are widely regarded as a versatile quantum engineering platform with abundant emergent phenomena [8-11]. Despite their promise, realizing TRS-protected transport in these materials has proven elusive [12], as opposed to simple binary compounds and 2D materials [13]. Although large Rashba SOC is indeed present at oxide interfaces due to inversion symmetry breaking [14], the TRS protection is often hampered by strong electron-electron interactions and high defect concentrations in oxides. Achieving TRS protection in correlated oxides is thus highly appealing in quantum engineering; however, it continues to pose significant experimental challenges.

A viable approach to this challenge is to explore quasi-1D oxide channels with strong SOC and enhanced sample quality. The reduced dimension in 1D will more effectively suppress backscattering paths with spin-momentum locking under TRS. Here in this work, we show evidence of TRS-protected transport on quasi-1D ferroelastic domain walls (FDWs) at ultraclean $LaAlO_3/SrTiO_3$ (LAO/STO) oxide interfaces. Notably, we report the observation of an anomalous light subband at modulation-doped 2D LAO/STO interfaces. The associated effective mass is remarkably light compared to the typical 3$d$ orbitals in STO-based electron systems. We investigate the induced quantum oscillations and back-gated multiband transport and reveal anomalous characteristics of this subband, including aperiodicity and susceptibility to the external magnetic field. These results are consistent with TRS-protected transport on quasi-1D FDWs with a nontrivial band topology. Finally, we explore the connection between superconductivity and quantum oscillations, suggesting the possible existence of a quantum critical point.

## INTERFACE ENGINEERING

The FDWs are ubiquitously formed in STO due to the structural phase transition at $T$=105 K. Spectroscopic and imaging studies have proven their existence and revealed the strain-tunable polarity [15,16], higher conductivity [17], and more robust superconductivity than the surrounding bulk [18]. It is intuitive to think that these 1D channels may have stronger SOC due to larger polarization on FDWs and thus be a natural place to study TRS-related physics. However, the impact of FDWs on electrical transport has rarely been observed in STO-based electron systems, possibly due to the presence of a large conducting background and oxygen vacancy ($V_O$) trapping on FDWs which reduce the SOC strength[19]. In addition, complex oxide interfaces are usually filled with defects compared to semiconductor heterostructures [20]. At the LAO/STO interface, oxygen vacancies ($V_O$), strontium vacancies ($V_{Sr}$), and ion intermixing constitute a comprehensive defect structure limiting mobilities beyond 10,000 cm²/Vs.

We design ($m+n$) uc LAO/STO interfaces to routinely achieve unprecedented mobilities over 10,000 cm²/Vs in pristine samples (Fig. 1(a)). Such sample configuration facilitates a modulation-doping mechanism [21,22]: the interfacial buffer LAO layer ($m=3$ uc) serves as a tunneling barrier with the major defects deliberately suppressed; the $V_O$ defects which serve as donors are solely confined in the top $n=2$ or 8 uc LAO layer through rapid annealing (RA) (See Sample growth and characterization and Fig. S1 in Supplementary Data).

Spectra from sum frequency generation (SFG) (Fig. 1(b)), a nonlinear optical process that is very sensitive to interfacial Ti-O bonds, show that the characteristic phonon mode for $V_{OS}$ (96-meV) is suppressed for (3+2) LAO/STO samples (red curve) while remaining prominent for traditionally grown (5+0) samples [23]. This result suggests the interface is virtually free of $V_{OS}$ in a (3+$n$) sample. Additional annular dark-field scanning transmission electron microscopy (ADF-STEM) shows that ion intermixing is also minimal (See Sample growth and characterization and Fig.S2 in Supplementary Data).

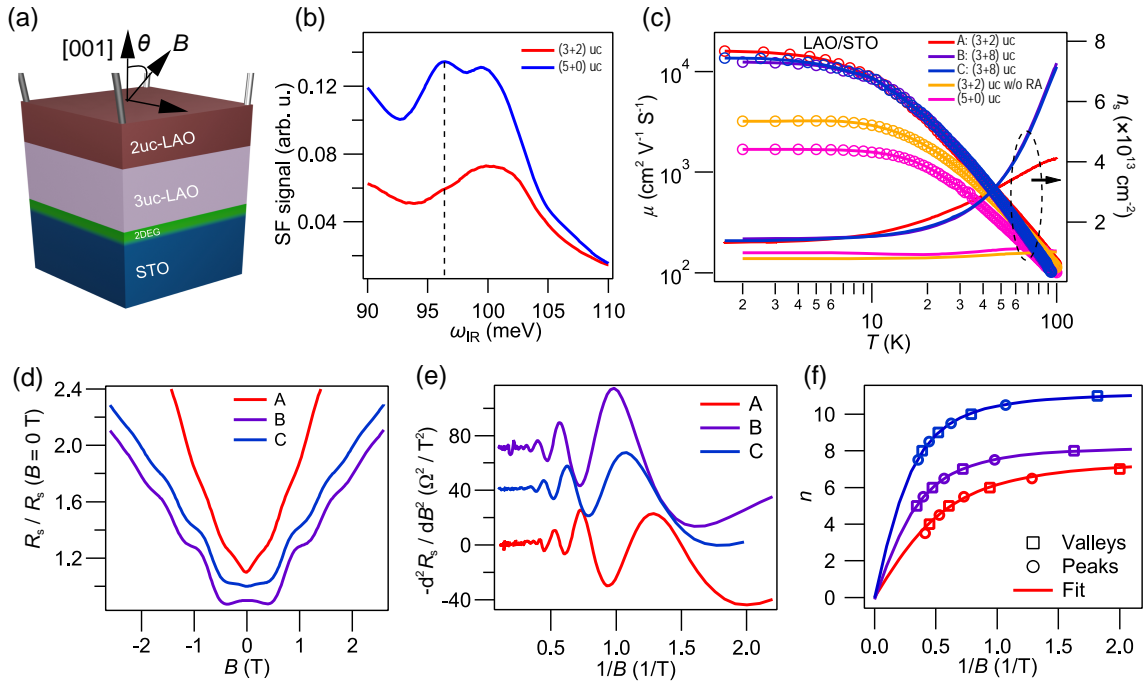

**Figure 1.** Interface design and anomalous quantum oscillations. (a) Schematic illustration of the sample layout. (b) SF spectra of a (5+0) sample (blue) and a (3+2) sample (red) suggest oxygen vacancies at the interface of (3+2) sample are suppressed. (c) Temperature dependence of carrier density and low-field ($B=0.5$ T) mobility (weighted average of all subbands). The modulation-doped samples (A, B, and C) show the highest mobilities. (d) The normalized sheet resistance of three samples (A, B, and C). (e) Aperiodic quantum oscillations in the $1/B$ axis. These oscillations start at a low field ~0.3 T with decreasing amplitudes at high fields. (f) Nonlinear Landau fan diagram. Solid lines are fittings to Eq. (1), open rectangles and circles are valleys (integer) and peaks (half integer) of $R_S$, respectively. The nonlinearity indicates the 1D to 3D transition of the transport on the FDWs.

As a result, the average low-field mobilities for as-grown (3+$n$) LAO/STO samples can easily reach above 10,000 cm$^2$/Vs, which is significantly higher than the traditional continuously grown (5+0) samples and (3+2) samples without RA (Fig. 1(c)). Further multiband analysis reveals that the mobilities of high mobility subband of (3+2) samples can reach over 46,000 cm$^2$/Vs.

## RESULTS

### Anomalous quantum oscillations

The quantum oscillations at the LAO/STO interface are rather complicated compared to simple 2DEGs on semiconductor heterostructures and 2D materials. The main discrepancy lies in inconsistent reports on the aperiodicity, assignment of the orbitals, and band degeneracies and crossings [24]. By studying nanowires fabricated by conductive atomic force microscope (cAFM), we previously proposed that channeled high mobility transport on FDW could reconcile all the known discrepancies at the 2D LAO/STO interface [25]. Specifically, the lateral confinement on quasi-1D FDW imposes additional energy on top of Landau quantization, which makes the energy ladder aperiodic (in the 1/$B$ axis) [26]. However, the reason why the carriers on FDWs and cAFM nanowires (FDW by nature) have enhanced mobilities and even perfect quantum ballistic transport remains a puzzle.

Figure 1(d) shows the normalized sheet resistances $\frac{R_\text{s}(B)}{R_\text{s}(0)}$ of three (3+$n$) LAO/STO samples (labeled as A, B, and C) at high back gate voltages $V_\text{bg}$ (~200 V). All the measurements were done at $T$=1.5 K in the van der Pauw geometry with 5 $mm$ ×5 $mm$ sample sizes. Although the profiles of magnetoresistance vary, they all show pronounced oscillations persistent to low magnetic fields ~0.3 T, especially for thicker (3+8) samples. The second derivative of sheet resistance $-\frac{\text{d}^2 R_\text{s}}{\text{d}B^2}$ reveals several anomalous features (Fig. 1(e)). First, these oscillations are all aperiodic on the 1/$B$ axis. Second, they appear primarily on low magnetic fields below 2 T. Third, the oscillation amplitudes rapidly decrease with increasing magnetic fields, in contrast to the standard Shubnikov-de Hass (SdH) oscillations [27]. Additionally, these oscillations are surprisingly reproducible in nearly all ~60 (3+$n$) samples grown.

The aperiodicity can be well explained by the magnetic depopulation effect on the quasi-1D FDWs. We analyze the oscillation in an extended Lifshitz-Kosevic (LK) framework by including geometric confinement [27]. Namely, the effective oscillation frequency is given by $\Omega = \sqrt{\omega_c^2 + \omega_y^2}$, where $\omega_c = eB/m_0$ and $\omega_y$ are cyclotron frequency and lateral confinement frequency of FDWs, respectively; $m_0$ is the cyclotron effective mass. As shown in Fig. 1(f), the positions of peaks and valleys in the oscillations can be fitted with a generalized Lifshitz–Onsager quantization rule with confinement renormalized to $\Omega$ and a dimensional phase shift,

$$n = \frac{\Omega_F}{\Omega} - \delta + \gamma, \quad (1)$$

where $n$ is the index of the magnetoelectric subband, $1/\Omega_F$ is the frequency for oscillations with respect to $1/\Omega$, $\gamma = \frac{1}{2} - \frac{\phi}{2\pi}$, $\phi$ is the Berry phase, and $\delta = 0$ ($\pm\frac{1}{8}$) is the phase shift for 2D (3D and 1D) dimensionality [28] (See Supplementary Data). The relation yields highly nonlinear $n \sim \frac{1}{B}$ curves in the low magnetic fields, as opposed to the linear dependence of Landau fan diagram in 2DEGs. This nonlinearity can be understood as a quasi-1D to 3D transition since $\omega_y$ is dominant only at low fields, while at high fields ($\omega_c \gg \omega_y$) the quasi-1D magnetoelectric subbands turn to regular Landau orbitals. We note the extraction of the Berry phase is unreliable due to the high sensitivity on the high field peak and valley positions, which have large uncertainties due to diminishing oscillation amplitudes.

**Angle dependence**

To further elucidate the dimensionality, we study the angle dependence of the anomalous quantum oscillation by rotating samples with respect to the magnetic field direction. As shown in Fig. 1(a) and Fig. 2, the oscillation amplitude gradually decreases as the field direction relative to the sample changes from out-of-plane ($\theta = 0°$) to in-plane ($\theta = 90°$). The peak and valley positions slightly shift at low magnetic fields while staying constant at high fields. This is consistent with quasi-1D to quasi-3D transition illustrated in the nonlinear Landau fan diagram (Fig. 1(f)). As the magnetic length $d_c = 2\sqrt{\frac{\hbar(2n+1)}{m_e^*\omega_c}}$ becomes smaller compared to the effective channel width $W = l_y^2\sqrt{\frac{2m_e^*\Omega_F}{\hbar}}$ on FDWs with increasing magnetic field, the quantum oscillations become increasingly 3D like in the high magnetic fields.

The angle dependence can rule out the possibility of $B$-periodic Sondheimer oscillations, whose oscillation period is inversely proportional to the thickness of the quasi-2DEG and should change with the sample rotation angle [29]. We also note that alternate explanations of the aperiodicity of quantum oscillations exist [30], which involve corrections from high-order magnetic response functions through the Gao-Niu quantization condition [31], or nonideal Dirac bands [32]. However, these scenarios mainly work for strong magnetic fields where high-order effects are nonnegligible, in contrast to the low-field aperiodic oscillation observed in this work. Finally, we emphasize that observing these oscillations does not necessitate sample-wide coherence; instead, it relies on uniform lateral confinement throughout the sample.

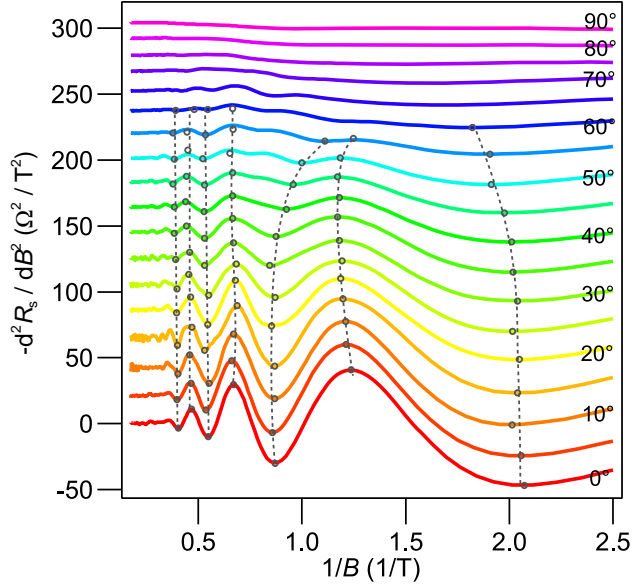

**Figure 2.** Angular dependence of the anomalous quantum oscillations. The second derivative of sheet resistance $-\frac{d^2 R_s}{dB^2}$ as a function of magnetic field angles $\theta$ ranges from $0°$ to $90°$. $\theta$ is the angle between the magnetic field direction and the [001] plane, as depicted in the schematic structure in Fig. 1(a). The data has been manually shifted for clarity. The dashed lines mark the evolution of oscillations with tilt angles, indicating the quasi-1D and 3D dimensionality at the low and high magnetic fields, respectively.

### Effective mass

The emergence of quantum oscillations at ultra-low magnetic fields suggests the associated carriers are very light, reminiscent of massless Dirac particles in clean graphene [33] and topological insulators [34], and light subbands in GaAs/AlGaAs quantum wells [35]. The effective mass can be measured from the temperature-dependent oscillation amplitudes in the LK equation $\frac{d^2 R_s}{dB^2} \propto \frac{\beta T}{\sinh(\beta T)}$, where $\beta = \frac{2\pi^2 k_B}{\hbar \Omega}$. This equation essentially describes the oscillation amplitudes decay with increasing temperature due to the thermal smearing of the electron distribution around the Fermi level, which is still valid for the extended LK model here (See Supplementary Data for details). Figure 3 shows the fitting of sample A data to the LK equation for oscillations for temperatures ranging from 1.6 K to 5 K. The extracted effective masses ($0.04 m_e \sim 0.08 m_e$) at different magnetic fields

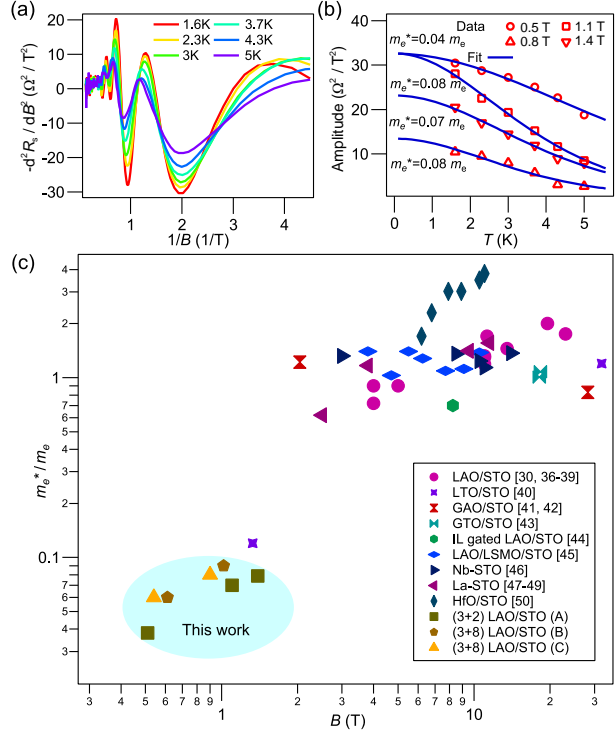

**Figure 3.** Effective mass. (a) Temperature dependent anomalous quantum oscillations of sample A. (b) Temperature dependent oscillation amplitude at different magnetic fields. Fitting to LK relation reveals the effective mass can be as small as 0.04$m_e$. (c) Survey of effective masses of STO based electron systems, which reveals unprecedently light electrons in this work.

are indeed extremely small, where $m_e$ is the bare electron mass.

Such a light subband is very surprising, provided the electronic structure of the LAO/STO interface has been extensively studied over the last two decades. In fact, as a prime platform for studying correlated physics, the effective mass of Ti $t_{2g}$ orbitals in STO-based electron systems is usually heavy ($0.6m_e < m_e^* < 4m_e$) [30,36-50], as shown in the survey of effective masses in literature in Fig. 3(c). The only exception is the ultra-thin LaTiO$_3$/STO (LTO/STO) interface reported by Veit *et al.*, where carriers with an effective mass of $0.12m_e$ were reported [40]. They proposed a phenomenological model by including a giant linear Rashba spin-orbit coupling (SOC) term with a Rashba parameter $\alpha_R = 1.8 \times 10^{-11}$ eV m, which significantly reduces the mass of inner hybridized $d_{xz+yz}$ orbitals. This scenario appears applicable to our work; however, $\alpha_R$ at the LAO/STO interface is typically one order smaller ($3.4 \times 10^{-12}$ eV m) [51], which is unlikely to support even lighter subbands.

**TRS-protected transport**

Considering the extensively studied 2D LAO/STO interface, the sole conceivable place for hosting such light carriers is the much less explored FDWs. It is well known that the band structure of the domain walls can be modified from the bulk materials due to local symmetry breaking in forms of band gap reduction, band bending, and metal-insulator transitions [52], as has been observed in materials including $BiFeO_3$ [53] and $Nd_2Ir_2O_7$ [54]. The FDWs in STO are polar due to local inversion symmetry breaking [55]. The associated polar field could, in principle, give rise to a much larger Rashba SOC than that of the interface, allowing novel electronic phases to emerge. For example, Yerzhakov et al. proposed that Majorana zero modes could appear on FDWs in STO by the interplay of strong Rashba SOC, ferroelectric polarization, and intrinsic superconductivity [55]. Fidkowski et al. proposed that the cAFM nanowire (FDW by nature) at the LAO/STO interface can be a "helical quantum wire," essentially an edge of a 2D quantum spin Hall insulator [56,57].

We find all the transport signatures are consistent with a linear quasi-1D Rashba model on FDWs. Although not as robust as topological protection, TRS protection is present due to the strong Rashba SOC on FDWs that enforces spin and momentum locking in quasi-1D and suppresses backscattering at low fields (See Supplementary Data and Fig. S6) [58]. The presence of $k$-linear Rashba SOC $H_{SO} = \alpha_R (\hat{n} \times \vec{k}) \cdot \vec{\sigma}$ leads to two energy bands $E_\pm(k) = \frac{\hbar^2 k^2}{2m_0} \pm \alpha_R |k|$ and creates two Fermi pockets, where $\hat{n}$ is the unit vector in the polarity direction, $\vec{\sigma}$ is the Pauli matrix and $m_0 = 0.6 m_e$ [49] is the lower limit of cyclotron mass of light bands experimentally observed in STO-based systems (Fig. 3). This scenario will predominantly happen in FDWs between Z and X (Y) domains, where $\hat{n}$ is perpendicular to $\vec{k}$ and X, Y and Z denote the tetragonal direction along crystal axes [15,55]. In the 2D/3D limit where a circular inner Fermi pocket can be approximated, we can estimate $\alpha_R$ using

$$m_e^* = m_0 \left(1 - \frac{\alpha_R}{\sqrt{\alpha_R^2 + 2\Omega_F \hbar^3/m_0}}\right), \quad (2)$$

For $m_e^* = 0.08 m_e$ at $B = 1.4$ T of sample A, a Rashba parameter $\alpha_R = 1.1 \times 10^{-10}$ eV m can be extracted. This value is on the same order of topological insulator $Bi_2Se_3$ ($4.0 \times 10^{-10}$ eV m) [59] and polar semiconductor BiTeI ($3.8 \times 10^{-10}$ eV m) [60], while being two orders larger than the LAO/STO 2D interface [14].

Generally, the amplitude of SdH oscillations increases with the magnetic field as the spacing between Landau levels widens. However, the observed low-field oscillations, both in resistance and conductance, surprisingly behave oppositely (See Fig. S5). This phenomenon can be understood as transport being protected by TRS at zero field in a quasi-1D Rashba wire. No backscattering channel is available since spins and momenta are tightly locked. Applying a magnetic field mixes spins and opens backscattering channels, eventually suppressing quantum oscillations by Anderson localization. This TRS-breaking process is commonly observed in the helical edge state of quantum spin Hall insulators in small magnetic fields [57,61]. In the meantime, high-field (up to 14 T) quantum oscillations are absent in most samples, possibly due to reduced mobilities at high fields (see Fig. S14 in Supplementary Data) after quenching the TRS-protected subband and enhanced mass for the outer Fermi pocket.

### Gate-tunable TRS-protected subband

Next, we show that the TRS-protected subband is gate-tunable and can be described by a multiband model with quantum corrections. Unlike semiconductor heterostructures, where electrical gating primarily adjusts the Fermi level, back gating at the LAO/STO interface mainly tunes the disorder and superconductivity [62,63]. Upon back gating ($-100$ V $< V_{bg} < 200$ V) at $T = 1.5$ K, the sheet resistance $R_s$ of sample A evolves from a simple parabolic shape without the quantum oscillations at $V_{bg} = -100$ V to a more complex dependence on the magnetic field at $V_{bg} > 0$ V (Fig. 4(a)). Correspondingly, Hall resistance $R_{xy}$ curves are relatively linear, and the Hall coefficient $R_H = \frac{R_{xy}}{B}$ stays constant at $V_{bg} = -100$ V (Fig. 4). With increasing $V_{bg}$, $R_{xy}$ curves become nonlinear with a characteristic "*S*" shape (Fig. 4(b)) and $R_H$ curves simultaneously develop a bell shape with fine features at low magnetic fields (Figs. 4(a), 4(b)). Two observations in $R_H$ are evident: 1) $R_H$ curve initially dips over a bell shape background ($V_{bg} < 0$ V), then flattens out at $V_{bg} = 0$ V and peaks sharply at $V_{bg} > 0$ V at low magnetic fields ($|B| < 1.5$ T); 2) Staircase features in $R_H$ emerge at high $V_{bg}$ values ($> \sim 100$ V) and low magnetic fields, which are clearly related to the quantum oscillations.

The nonlinear Hall effect at the LAO/STO interface has been predominantly explained by the two-band model in the literature, in which the characteristic "*S*" shape in $R_{xy}$ and the bell shape in $R_H$ are caused by a second high-mobility electron band [64,65]. Although not routinely studied, the dip feature inside the bell shape in $R_H$ curve most naturally suggests the emergence of hole carriers. This possibility is immediately dropped in literature since the TiO$_2$-terminated LAO/STO interface is *n*-type. However, with the presence of the Dirac point in the proposed linear Rashba model, this hole picture is still possible, and the dip-to-peak transition in $R_H$ can be understood as the hole-to-electron transition. Alternatively, Gunkel et al. added an anomalous Hall effect (AHE) term originating from interfacial magnetism to the two-band model to explain the dip feature. This combined AHE model works well with NdGaO$_3$/STO and defect-tuned LAO/STO interfaces, which are magnetic and low mobility [65,66]. While magnetism at LAO/STO interfaces is still highly controversial [67], our ultra-clean LAO/STO samples are presumably non-magnetic due to the absence of interfacial oxygen vacancies. In addition, no magnetic hysteresis is observed in all $R_s$ and $R_{xy}$ curves (See Supplementary Data and Fig. S8).

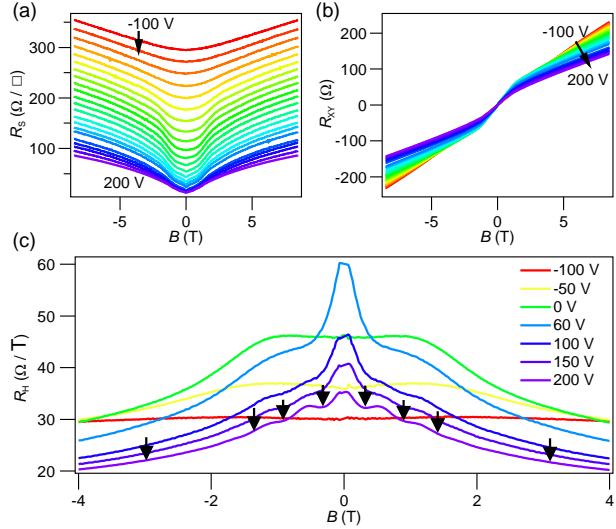

**Figure 4.** Gate-tunable transport of Sample A. (a) Sheet resistance $R_s$ as $V_{bg}$ sweeps from $-100$ V to 200 V. $R_s$ evolves from a simple parabolic shape to a more complex dependence on the magnetic field. (b) Simultaneously measured $R_{xy}$ curves become nonlinear with a characteristic "S" shape, suggesting multiband transport. (c) Hall coefficient $R_H$ at different back gate voltages $-100$ V $< V_{bg} <$ 200 V. The profile of $R_H$ turns from constant to a bell shape ($-4$ T $< B <$ 4 T), with a dip-to-peak transition in low field ($-1.5$ T $< B <$ 1.5 T) at $V_{bg} > -50$ V. The black arrows mark the quantum oscillations at $V_{bg} = 200$ V.

Nevertheless, we check the applicability of the hole picture and AHE picture by fitting the transport data with a 3-band model, which includes an additional band to reflect hole-to-electron transition or AHE contribution on the basis of the classical two-band model. As a result, neither model can faithfully fit the data across the dip-to-peak transition in $R_H$ (Fig. 5(a)) and yields unrealistic fitting parameters, as is evident in the goodness of fit using the root-mean-square-error (RMSE) (See Fig. S9 and Fig. S10 in Supplementary Data). Furthermore, our gating data does not fall under the universal scaling law at LAO/STO interfaces observed by Joshua *et al.* [64], which states that a critical carrier density ($\sim 1.6 \times 10^{13}$ cm$^{-2}$) exists at the Lifshitz transition from $d_{xy}$ to hybrid $d_{xz+yz}$ bands (See Fig. S11). These deviations suggest that our ultra-clean samples cannot be described by the classical multiband model with a rigid band structure.

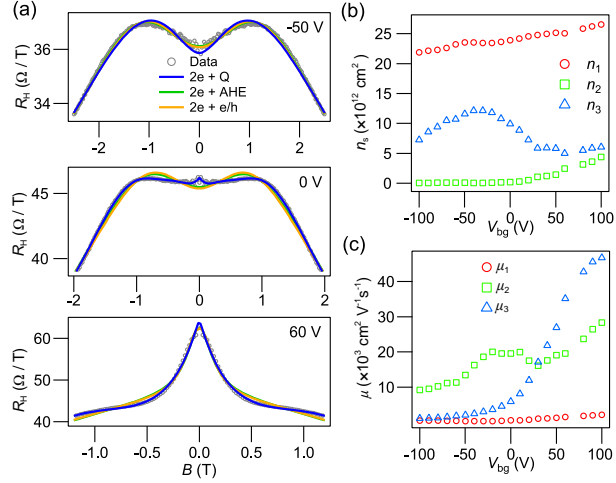

**Figure 4.** Quantum-corrected multiband fitting. (a) Fitting with various versions of the multiband models. It is obvious that the quantum-corrected multiband model yields the best fitting. 2e+Q: Quantum corrected multiband model; 2e+AHE: 2-band model with AHE; 2e+e/h: 2-band model with electron to hole transition. Carrier densities (b) and mobilities (c) of the three subbands at $B=0$ T, respectively, extracted from the 2e+Q model.

All the features in the back-gating data are consistent with the proposed quasi-1D Rashba model and multiband transport with quantum corrections. In the classical multiband model, carrier densities and mobilities are independent of the magnetic field. However, the TRS-protected transport on FDWs is clearly field-dependent. It manifests itself as the mobility of the associated subband sharply decreases in the TRS-breaking process caused by the magnetic field. We rewrite the three-band model to

$$R_H = -\frac{1}{e} \frac{\sum_j \frac{n_j \mu_j^2}{1+\mu_j^2 B^2}}{\left(\sum_j \frac{n_j \mu_j}{1+\mu_j^2 B^2}\right)^2 + \left(\sum_j \frac{n_j \mu_j^2}{1+\mu_j^2 B^2}\right)^2 B^2}, \quad (3)$$

where the $n_j$ and $\mu_j$ ($j=1,2,3$) are the carrier density and mobility of each subband, $\mu_3(B) = \mu_3^0 \exp(-\frac{B}{B_c})$ is the mobility of the subband on the FDWs corrected by an exponential dependence of magnetic field, with $\mu_3^0$ the zero-field mobility and $B_c$ the characteristic field (See Supplementary Data). This model provides an excellent fit to the data by capturing all the details of the dip-to-peak transition. Figures 5(b) and 5(c) show the extracted $n_j$ and $\mu_j$, which reveal a quantum-corrected subband ($n_3$, $\mu_3$) with negative electronic compressibility and rapid onset of mobility from ~6,000 cm²/Vs ($V_{bg} = 0$ V) to over 46,000 cm²/Vs ($V_{bg} = 100$ V).

## DISCUSSION

All the experimental evidence, including the aperiodic quantum oscillations at the ultra-low magnetic fields, extremely small effective mass, suppression of oscillation at high magnetic fields, and quantum-corrected multiband transport, are consistent with the TRS-protected transport on FDWs with a giant Rashba SOC. Under this framework, we next discuss possible insights into the complex LAO/STO interface.

We can understand why this TRS-protected state has not been explicitly observed over the past two decades. It is well-known that FDWs are prone to trap oxygen vacancies [19], which compensate the polar field [20] and subsequently reduce the Rashba SOC. In addition, oxygen vacancies increase disorder, which easily causes backscattering in the quasi-1D dimension. The oxygen vacancies are intentionally suppressed in our ultra-clean LAO/STO samples, as revealed by SF spectra and transport measurements. This could revive the strong Rashba SOC on FDWs and lead to the observed anomalous quantum oscillations.

A related question is whether the weak antilocalization (WAL) effect is relevant, which manifests as a sharp dip in low field $R_s$ and is frequently used to study interfacial Rashba SOC [14]. There are no definitive signatures of WAL in the four samples reported here, either nonexistent or too small compared to the quantum-corrected multiband transport (Fig. 4). We point out that WAL is a 2D effect that should be suppressed in quasi-1D FDWs due to the absence of interfering scattering paths.

The strongly ionic nature of the complex oxide interface leads to an intricate set of subbands displaced vertically in real space. In principle, including more bands in the multiband model will describe the interface more precisely [68]; however, the three-band model can at least qualitatively tell the subband characteristics. It is plausible to recognize the quantum-corrected band ($n_3, \mu_3$) as the same subband of anomalous quantum oscillations due to their inherent sensitivity to the magnetic field and back gating. Namely, back gating tunes disorder level and Rashba SOC strength and gives rise to the enhanced mobility and anomalous quantum oscillation at high $V_{bg}$. Meanwhile, subband ($n_1, \mu_1$) with high density and low mobility can be attributed to carriers close to the interface or the outer Fermi pocket on FDWs, both with a larger mass and higher density. Subband ($n_2, \mu_2$) likely originates from high mobility interfacial electrons, which quickly populate at $V_{bg} > \sim 30$ V.

Finally, it is intriguing to know the relationship between superconductivity and anomalous quantum oscillations for a correlated interface. Superconductivity in STO is often related to the ferroelectric mode [69,70], with electron pairing mediated by polar instability or ferroelastic waves on FDWs [71]. Indeed, superconducting $T_c$ is found to be higher on FDWs [18]. We cool down samples A to mK temperatures and extract a phase diagram as a function of temperature and $V_{bg}$. Interestingly, superconductivity persists up to the $V_{bg}$ values at which anomalous quantum oscillations start to appear (within ~50 V due to sample inhomogeneity and gate uncertainty), as shown in Fig. S12 and Fig. S13. This coincidence suggests the possible existence of a quantum critical point that marks the transition between superconductivity and quantum oscillations. Moreover, while $n_1$ is mostly constant and $n_2$ is not large enough to support superconductivity in the range of $-40\,\text{V} < V_{\text{bg}} < 50\,\text{V}$, the carrier density $n_3$ of the quantum-corrected band $(n_3, \mu_3)$ vary significantly (over 50%), which matches the profile of the superconducting dome. This observation points to a direct connection between superconductivity and the FDWs.

## CONCLUSIONS

In summary, our experiments uncover an anomalous light subband at the correlated LAO/STO interfaces. All the experimental evidence points to a TRS-protected state on quasi-1D FDWs, with a Rashba SOC strength in the same order as typical topological insulators. Our work adds the missing TRS-protected physics to the rich set of emergent phenomena, paving the way to quantum engineering of novel states of matter at the oxide interfaces.

## ACKNOWLEDGMENTS

We thank Zhenyu Zhang, Qian Niu, Zhengfei Wang, Wenguang Zhu, and Zhenyu Ding for their helpful discussions.

## AUTHOR CONTRIBUTIONS

GC conceived the ideas; MH and QX investigated and carried out transport experiments; ZQ, CM, DL, MH and ZL grew and fabricated samples; MH, QX, GC and YG performed data analysis; CKD and LS performed DFT calculations; KC and PY performed ADF-STEM characterization; WTL and XL performed SFG measurements; MH visualized images; GC, MH, YG, QX wrote paper with input from all co-authors.

## DATA AVAILABILITY

The data supporting the findings of this study are available from the corresponding author upon reasonable request.

SUPPLEMENTARY MATERIALS

## SAMPLE GROWTH AND CHARACTERIZATION

1. Sample growth

LAO/STO samples were epitaxially grown by pulsed laser deposition (PLD) equipped with a reflection high-energy electron diffraction (RHEED) system. Before growth, (100) STO substrates were treated by buffered HF and annealed in an oxygen environment at 950 °C to achieve an atomically smooth surface. The laser spot has a flat-top profile, which is crucial to achieving the desired results. The laser fluence is 0.9 J/cm$^2$ with 1 Hz repetition frequency. The sample growth was broken into five stages (Fig. S1):

1) Pre-annealing. STO substrate is heated up to 650 °C by a laser in $P_{O_2} = 100$ Torr to remove contaminants at the surface.

2) Interfacial LAO growth. The interfacial buffer LAO layer (*m*=3 uc) is deposited at 550 °C with a low oxygen partial pressure $P_{O_2} = 5 \times 10^{-6}$ Torr to introduce abundant V$_O$s, which compensates for the built-in potential $\phi_{LAO}$ arising from polar discontinuity and reduce ion intermixing (driven by $\phi_{LAO}$) and V$_{Sr}$ formation at the interface.

3) Intermediate annealing. These interfacial V$_O$s are subsequently removed by annealing in high $P_{O_2} = 200$ Torr (for 1~3 hours) at a lower temperature *T*=430 °C, at which ion intermixing is still mitigated due to much lower diffusion coefficients of metal ions than that of oxygen ions. This procedure creates virtually defect-free LAO/STO interfaces.

4) Top LAO growth. The top *n*=2 or 5 uc LAO layer is then grown at the annealing temperature (430 °C) with a low $P_{O_2} = 5 \times 10^{-6}$ Torr to generate V$_O$ donors, which transfer electrons to the interface. RHEED oscillations are observed in both procedures (Fig. S1). Such (*m*+*n*) configuration essentially enables a modulation doping charge transfer process.

5) Rapid annealing (RA). Finally, the (*m*+*n*) LAO/STO sample is rapidly cooled to room temperature within 30 seconds in an oxygen environment (tuned to several Torr to control the carrier densities) once the laser is turned off and is immediately transferred to the nitrogen-filled load lock to provide further cooling. This RA procedure is vital to prevent the migration of V$_O$s to the interface and degrade the interface quality.

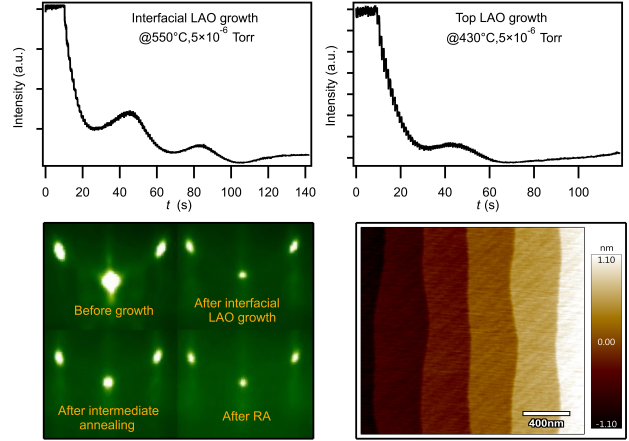

**Fig.S1. (*m+n*) LAO/STO sample growth.** (A) and (B) show RHEED oscillations for the interfacial m uc and top n uc LAO growth. (C) RHEED patterns at various growth stages. (D) AFM image shows an atomically flat surface after growth.

The (5+0) samples were continuously grown at the same growth conditions, except *m*=5 and without the top LAO growth stage. This is the conventional method for growing LAO/STO samples, which serve as control samples in this work.

## 2. Sum Frequency Generation (SFG) Spectroscopy

We utilize SFG phonon spectroscopy to probe the interfacial oxygen vacancies in (3+2) uc and (5+0) uc LAO/STO samples[S1]. Broadband infrared and narrowband pulses, generated by a 35-fs pulsed laser amplifier (Spitfire ACE, Spectra Physics) of 800 nm wavelength at 2 kHz repetition rate, are overlapped at the sample surface after pumping an optical parameter amplifier followed by a difference frequency generation stage (TOPAS-C, Spectra-Physics). The resultant SFG signal is very sensitive to inversion symmetry breaking, an ideal tool for probing interfacial phonons in STO. We previously reported that the 96-meV resonance peak is characteristic of $V_O$s at the interface. This peak is suppressed in (3+2) uc samples, indicating $V_O$ is significantly less in these samples than in continuously grown (5+0) samples.

## 3. ADF-STEM

ADF-STEM is performed by the Grand ARM 300F (JEOL) with an aberration corrector operated at 300 kV. As shown in Fig. S2, the high-resolution mapping image shows the atomically abrupt interface of (3+2) uc LAO/STO interface, with no sign of ion intermixing. The detailed structure of the top 2 uc is less sharp, suggesting the deficiency of oxygen atoms. We also detect oxygen deficiency along the top atomic plane and $V_O$s are obtained expectedly. These characterizations suggest the modulation doping mechanism in (3+2) uc LAO/STO heterostructure.

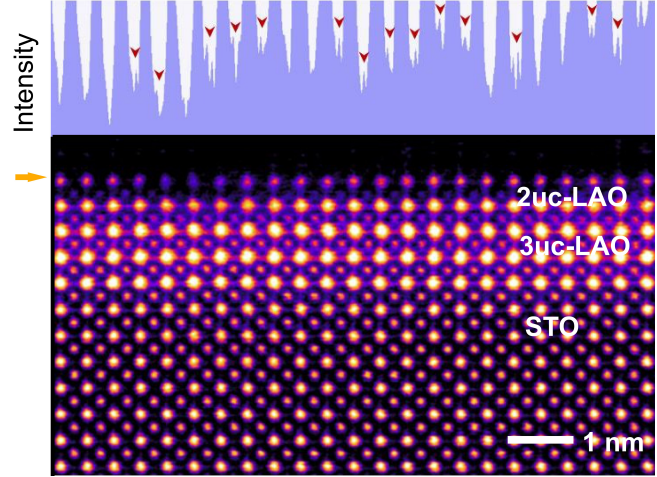

**Fig. S2. ADF-STEM imaging of sample A.** Red arrows in the top panel mark oxygen peaks (with some columns missing due to oxygen deficiency) along the atomic plane indicated by the yellow arrow.

## ENERGY DIAGRAM and Calculation details

Density functional theory (DFT) calculations were used to analyze the defect formation in the LAO layer in such a growth strategy. The formation energy $\Delta H(\varepsilon_F(x), X^q)$ of defect $X^q$ in LAO depends on the distance $x$ between the defect and the LAO/STO interface, since the Fermi level relative to local valence band maximum (VBM) $\varepsilon_F(x)$ is different due to the polarization electric field within LAO. As shown in Fig. S3(a), in $O_2$ poor environment, the $\varepsilon_F(3\ uc)$ is ~1.25 eV and the formation energy of $V_O$s is close to zero. The formation of antisite defects $Ti_{Al}^+$ is also suppressed in $O_2$ poor environment, so the major defect in LAO is $V_O$ during the first interfacial 3 uc LAO growth, consistent with the previous calculations [S2]. The migration energy of $Ti^{4+}$ (~8.3 eV) is much higher than $O^{2-}$ (~0.5 eV) and is still high (~4.2 eV) with the assistance of $V_{Sr}$, so the annealing at low temperature and rich $O_2$ atmosphere can effectively eliminate $V_O$s without activating ion intermixing at the interface. In addition, the migration of $Sr^{2+}$ is also slow with ~3.7 eV migration energy and contributes less compensating field. In the first 3 uc LAO after annealing, the electric field remains high, and the Fermi level $E_F$ is slightly lower than the conduction band minimum (CBM) (Fig. S3(b)).

Finally, the fast quenching of the top 2 uc LAO layers makes them rich in intrinsic defects, such as $V_{La}$, $V_{Al}$ and $V_O$, which will compensate the field as shown in Fig. S3(b). Meanwhile, their relative amount can be tuned by growth oxygen partial pressure. In $O_2$ poor environment, more $V_O$ defects will be formed, which donate more electrons to the CBM of STO to form 2DEG. Therefore, a clean interface is expected with such a sample growth strategy.

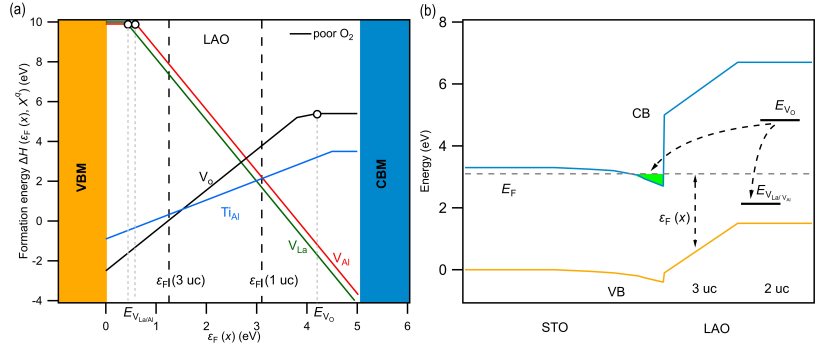

**Fig. S3. Energy diagrams calculated by DFT.** (a) The formation energy diagram of the intrinsic defects and the $Ti_{Al}$ defect in LAO, where the solid polylines are calculated with the $O_2$-poor ($\Delta\mu_O = -2$ eV) growth condition. The vertical dashed line represents the Fermi levels relative to the VBM at the 1 uc and 3 uc of the LAO slab. The open circles indicate the defect charge transition energy levels $E_{V_{La/Al}}$ and $E_{V_O}$. (b) The band alignment schematic of 5 uc LAO slab on STO substrate, where the horizontal dashed line represents the Fermi level of the sample.

All calculations were performed using DFT and the plane-wave projector-augmented wave method as implemented in the VASP code. An energy cutoff of 520 eV and the PBEsol exchange correlation functional was adopted to relax the structures. HSE06 hybrid functional was used to calculate the energy on top of these relaxed structures for better describing the defect level and band gap in the formation energy calculation. The Brillouin zone was sampled by $6 \times 6 \times 6$ and $1 \times 1 \times 1$ $k$-point mesh centered on $\Gamma$ point for primitive cell of 5 atoms and supercell of 135 atoms, respectively. The atomic forces were relaxed to be less than 0.01 eV/Å. The calculated lattice parameters of STO and LAO are 3.895 Å and 3.771 Å.

## TRANSPORT MEASUREMENT

We use reverse-field reciprocity to avoid sweeping the field, which is applicable as far as the sample obeys the Onsager reciprocal relation (i.e. ohmic), and does not show magnetic hysteresis [S3]. We also adopt a spinning current Hall measurement method in a van der Pauw (vdP) geometry to achieve fast and precise measurement. The method is frequently used in commercial Hall sensors to achieve nanotesla sensitivity [S4].

Specifically, 4 aluminum wires are punched at corners to contact the interface by wire bonding in typical $5\,mm \times 5\,mm$ samples, as shown in Fig. S4. Eight measurement configurations, labeled from A to H in Table S1, are quickly permutated in a small magnetic field ($B$=0.5 T) using a matrix switch. In each configuration, an *I-V* curve is measured to extract the resistance $R_{ij,mn}$, which represents the resistance measured by sourcing current $I_{ij}$ from contact *i* to *j*, and measuring the voltage difference $V_{mn}$ between contact *m* and *n*. The linearity of each *I-V* curve is also used as a sanity check for the quality of ohmic contact of wire bonds at any conditions, which we found extremely important in correctly interpreting experiment data.

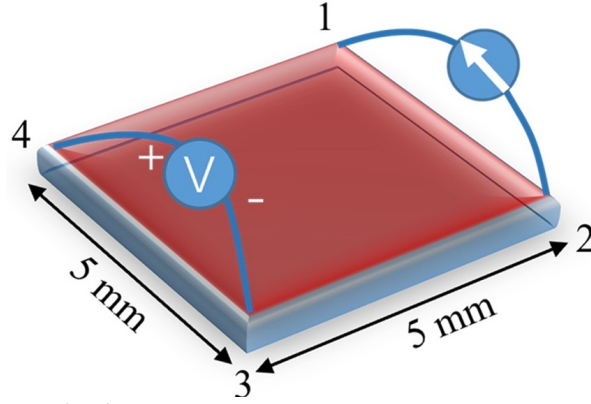

**Fig. S4. Configuration A of the spinning current Hall measurement.** Contact 1 and 2 are connected to source current. Contact 3 and 4 are connected to a voltmeter.

For longitudinal configurations A to D, the common false signals in longitudinal voltage measurement are thermoelectric voltage $V_{th}^{mn}$ originating from the temperature difference between contacts $m$ and $n$, and misalignment voltage originating from the crosstalk between Hall voltage and longitudinal voltage due to geometrical misalignment of the contacts. The thermoelectric voltage can be easily removed by differentiating the $I$-$V$ curve since it is mostly constant in a single $I$-$V$ curve measurement. As a result, $R_{12,43}$ for configuration A can be written as

$$R_{12,43} = \frac{dV_{43}}{dI_{12}} = R_a + c_1 R_H B, \tag{S1}$$

where $R_a$ is one of the two resistance values used in the van der Pauw formula, $R_H$ is the Hall coefficient, $B$ is the magnetic flux density and $c_1$ is a small constant due to contact misalignment. According to reverse-field reciprocity, interchanging the current and voltage measurements (i.e., from configuration A to configuration C) effectively flips the magnetic field. Then we have

$$R_{34,21} = \frac{dV_{21}}{dI_{34}} = R_a - c_1 R_H B, \tag{S2}$$

Averaging these two resistance values yields

$$R_a = \frac{1}{2}(R_{12,43} + R_{34,21}), \tag{S3}$$

Similarly, we have the other resistance value $R_b$ in van der Pauw measurement

$$R_b = \frac{1}{2}(R_{23,14} + R_{41,32}), \tag{S4}$$

The sheet resistance $R_s$ is then calculated by solving the van der Pauw equation

$$\exp\left(-\frac{\pi R_a}{R_s}\right) + \exp\left(-\frac{\pi R_b}{R_s}\right) = 1, \tag{S4}$$

Meanwhile, configurations E to H are applied to measure the Hall coefficient. The contribution of longitudinal voltage due to the misalignment of contact leads is eliminated by reverse-field reciprocity and averaging the four values:

$$R_H B = \frac{1}{4}(R_{13,42} + R_{24,13} + R_{13,42} + R_{13,42}), \tag{S5}$$

The carrier density $n_e$ and low-field Hall mobility $\mu_H$ are then given by

$$n_e = \frac{1}{eR_H}, \tag{S6}$$

$$\mu_H = \frac{1}{n_e e R_s}, \tag{S7}$$

where $e$ is the electron charge. All the configuration information is listed in Table S1.

**Table S1. Configuration setup of the spinning current Hall measurement.** For every configuration, current enters the sample at contact "$I_+$" and leaves at contact "$I_-$", voltage difference is measured between contact "$V_+$" and "$V_-$". The measured voltage consists of 3 contributions: Hall voltage, longitudinal voltage and thermoelectric voltage. For vdP configurations (A, B, C and D), the small Hall voltage part ($\pm c_{1,2} R_H BI$, $|c_{1,2}| \ll 1$) is due to misalignment. For Hall configurations (E, F, G and H), misalignment leads to a small longitudinal voltage contribution ($\pm \alpha R_s I$, $|\alpha| \ll 1$). Thermoelectric voltage $V_{th}$ is eliminated by calculating the slope of *I-V* curve, which is defined as the measured resistance of a single *I-V* curve measurement.

| Configuration | \multicolumn{4}{c}{Contact name} | \multicolumn{3}{c}{Measured voltage $V$} | $dV/dI$ |
|---|---|---|---|---|---|---|---|---|
| | $I_+$ | $I_-$ | $V_+$ | $V_-$ | Hall voltage | Longitudinal voltage | Thermoelectric voltage | |
| A | 1 | 2 | 4 | 3 | $c_1 R_H BI$ | $R_a I$ | $V_{th}^{43}$ | $R_{12,43} = R_a + c_1 R_H B$ |
| B | 2 | 3 | 1 | 4 | $c_2 R_H BI$ | $R_b I$ | $V_{th}^{14}$ | $R_{23,14} = R_b + c_2 R_H B$ |
| C | 3 | 4 | 2 | 1 | $-c_1 R_H BI$ | $R_a I$ | $V_{th}^{21}$ | $R_{34,21} = R_a - c_1 R_H B$ |
| D | 4 | 1 | 3 | 2 | $-c_2 R_H BI$ | $R_b I$ | $V_{th}^{32}$ | $R_{41,32} = R_b - c_2 R_H B$ |
| E | 1 | 3 | 4 | 2 | $R_H BI$ | $\alpha R_s I$ | $V_{th}^{42}$ | $R_{13,42} = R_H B + \alpha R_s$ |
| F | 2 | 4 | 1 | 3 | $R_H BI$ | $-\alpha R_s I$ | $V_{th}^{13}$ | $R_{24,13} = R_H B - \alpha R_s$ |
| G | 3 | 1 | 2 | 4 | $R_H BI$ | $\alpha R_s I$ | $-V_{th}^{42}$ | $R_{13,42} = R_H B + \alpha R_s$ |
| H | 4 | 2 | 3 | 1 | $R_H BI$ | $-\alpha R_s I$ | $-V_{th}^{13}$ | $R_{13,42} = R_H B - \alpha R_s$ |

# LANDAU FAN, RASHBA MODEL AND REDUCED EFFECTIVE MASS

We plot the conductance of each sample as a function of $1/B$ with a smooth background removed (Fig. S5). The conductance oscillations show similar characteristics of resistance oscillations shown in Fig. 2(b), which are opposite to typical SdH oscillations.

To capture all these characteristics mentioned in the main text, we start from the Lifshitz–Onsager quantization in a typical high-mobility 2DEG,

$$n = \frac{hS_F}{4\pi^2 eB} - \gamma, \tag{S8}$$

where $S_F$ is the extremal cross-sectional area of the Fermi surface, $\gamma = \frac{1}{2} - \frac{\phi}{2\pi}$, and $\phi$ is the Berry phase [S5]. When it comes to a quasi-1D system where the effective width $W$ in real space is comparable to the magnetic length $d_c$, extra lateral electrostatic confinement needs to be considered in addition to the confinement caused by the applied perpendicular magnetic field. For free electrons, we have the Hamiltonian (See reference [S6])

$$\hat{H} = \frac{\hat{p}_y^2}{2m^*} + \frac{m^* \Omega^2}{2} y^2 + \frac{\hbar^2 k_x^2}{2m^*}, \tag{S9}$$

where $\Omega = (\omega_c^2 + \omega_y^2)^{\frac{1}{2}}$, $\omega_y = \frac{\hbar}{m_0 l_y^2}$, $m^*$ is the effective cyclotron mass, $l_y$ is the characteristic width of the lateral confinement potential and $m_0$ is the cyclotron mass. Therefore, the quasi-1D nature renormalizes the strength of the magnetic field and adds an additional $k_x$ dependence in addition to the harmonic oscillator Hamiltonian. The latter is similar to the Landau level in 3D where an additional $k_z$ dependence also exists. So the corresponding energy spectrum $\epsilon_n$ is given by

$$\epsilon_n(k_x) = \hbar\Omega\left(n + \frac{1}{2}\right) + \frac{\hbar^2 k_x^2}{2m^*}\frac{\omega_y^2}{\Omega^2}, \tag{S10}$$

which is a series of vertically displaced parabolic magnetoelectric subbands and has a similar form to the 3D case.

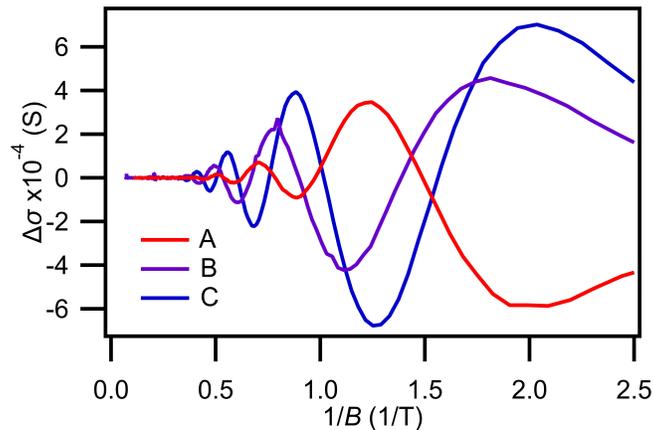

**Fig. S5. Conductance plot of quantum oscillations in Sample A, B and C.** The oscillation is aperiodic in $1/B$ axis, and the oscillation amplitude decreases with increasing magnetic field.

For quasi-1D electrons with a general energy-momentum dependence, we adopt the same treatment as before, i.e., the confining potential modifies the magnetic field and an additional $k_x$ dependence is present. In that case, Eq. (S8) becomes

$$n = \frac{hS(\varepsilon, k_x)/m^*}{4\pi^2\Omega} - \gamma, \tag{S11}$$

The resulting spectrum is therefore $\varepsilon = \varepsilon(n, k_x)$.

Following the treatment of SdH oscillations in the LK framework in reference [S7], we extend the analysis and find that

$$\rho = \rho_0\left[1 + A(\Omega, T)\cos 2\pi\left(\frac{\Omega_F}{\Omega} - \delta + \gamma\right)\right], \tag{S12}$$

where $A(\Omega, T) = \frac{\beta T}{\sinh(\beta T)} e^{-\beta T_D}$, with $\beta = 2\pi^2 k_B/\hbar\Omega$, $k_B$ is the Boltzmann constant, $T_D$ is Dingle temperature, $1/\Omega_F$ is the frequency for oscillations with respect to $1/\Omega$, and $\delta$ is an additional phase shift determined by the dimensionality, i.e., $\delta = 0$ for the 2D case and $\delta = \pm 1/8$ for the 3D or quasi-1D case. The positions of peaks and valleys in the oscillations can thus be described in Eq. (1) in the main text.

The energy spacing of a quasi-1D system in a magnetic field is described by $\hbar\Omega$, which no longer gives a linear relation between $n$ and $1/B$ in low fields as in 2D. With increasing magnetic fields, the energy spacing $\hbar\Omega \to \hbar\omega_c$ as Landau level spacing $\hbar\omega_c$ gradually dominates over the lateral electrostatic confinement. Meanwhile, the Landau level occupation turns to approximate linear in a pure 2D system, so $\gamma$ takes the formula of $\gamma = \frac{1}{2} - \frac{\phi}{2\pi}$.

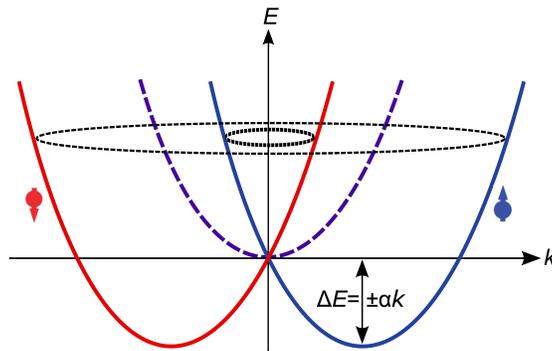

**Fig. S6. Rashba model on FDWs.** Two Fermi pockets emerge with the inclusion of the linear Rashba SOC in the Hamiltonian.

In brief, the generalized Lifshitz–Onsager quantization rule can be described as equation S11. By analyzing the $n \sim 1/B$ curves (Landau fan diagram) of the quantum oscillations, we can obtain values of $\Omega_F$, $l_y$, $d_c$ and $W$ of sample A~C listed in Table S2.

The inversion symmetry breaking and the polarity on FDWs suggest the possibility of a large Rashba SOC, characterized by a $k$-linear Rashba model

$$\hat{H} = \frac{\hbar^2 k^2}{2m_e^*} \pm \alpha_R |k|, \tag{S13}$$

As shown in Fig. S6, two Fermi pockets are emerging. In the 2D/3D limit for a fixed Fermi surface, the effective mass is given by

$$m_e^* = \frac{\hbar^2}{2\pi} \frac{\partial S_F}{\partial \epsilon_F}, \tag{S14}$$

where $S_F \sim \pi k^2$ is the area of the enclosed Fermi pocket, and the partial derivative is evaluated at the Fermi energy. In this case, we have the same explicit form for effective mass related to the inner fermi pocket that is responsible for low-field quantum oscillations

$$m_e^* = m_0 \left(1 - \frac{\alpha_R}{\sqrt{\alpha_R^2 + 2\Omega_F \hbar^3/m_0}}\right), \tag{S14}$$

At a given magnetic field, the temperature dependence of the oscillation amplitude can be described by the extended Lifshitz-Kosevich model: $d^2 R_s/dB^2 \sim \beta T/\sinh(\beta T)$, with $\beta = 2\pi^2 k_B/\hbar\Omega$. We fit the amplitudes of the second derivative of oscillations (from those we can reliably measure) to the model and show the results in Fig. S7 in addition to Fig. 4(b), which yields effective masses less than $0.1 m_e$. These exceptional light electrons are possibly arising from the inner Fermi surface produced from the bottom of the $d_{xy}$ band on FDWs.

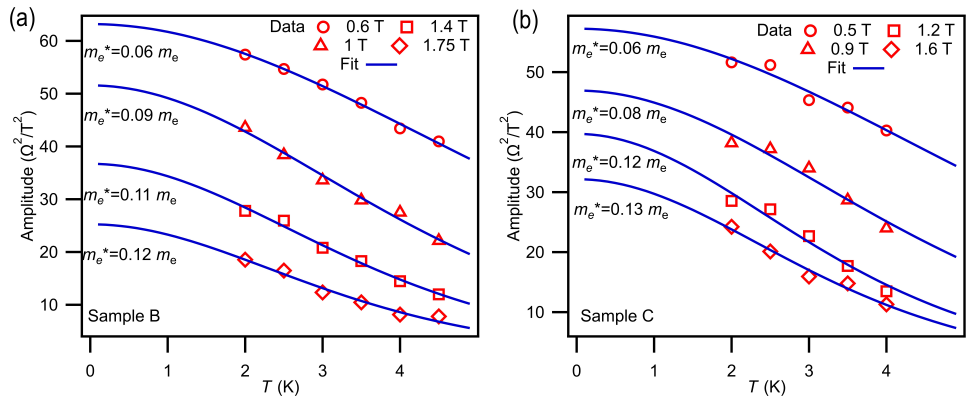

**Fig. S7. Effective mass extraction for Sample B and C.** Extended L-K model fitting of temperature-dependent oscillations in $d^2 R_s/dB^2$ for Sample B (a) and Sample C (b).

**Table S2.** Subband parameters extracted from the analysis of the generalized Lifshitz–Onsager quantization rule and Rashba model.

| Sample | $m_e^*$ ($m_e$) | $\Omega_F$ ($s^{-1}$) | $l_y$ (nm) | $\alpha_R$ (eVm) | $d_c$ (nm) | $W$ (nm) |
|---|---|---|---|---|---|---|
| A | 0.04@0.5 T<br>0.07@1.1 T<br>0.08@1.4 T | $2.4 \times 10^{13}$ | 21 | $1.1 \times 10^{-10}$@1.4 T | 218@0.5 T<br>104@2.2 T | 83 |
| B | 0.06@0.6 T<br>0.09@1.0 T | $2.4 \times 10^{13}$ | 17 | $1.3 \times 10^{-10}$@1.0 T | 220@0.6 T<br>100@2.9 T | 70 |
| C | 0.06@0.5 T<br>0.08@0.9 T | $2.8 \times 10^{13}$ | 17 | $1.4 \times 10^{-10}$@0.9 T | 262@0.5 T<br>115@2.6 T | 74 |

## MULTIBAND TRANSPORT

As $V_{bg}$ sweeps from -100 V to 200 V, the sheet resistance $R_s$ of sample A evolves from a simple parabolic shape to a more complex dependence on the magnetic field. Meanwhile, $R_{xy}$ curves become nonlinear with a characteristic "S" shape (Fig. 4), suggesting multiband transport. Multiband transport is frequently applied to oxide interfaces where multiple electron orbitals are responsible for the transport. This model assumes that the mobilities of carriers do not change with the magnetic field and usually yields relatively good fitting for low mobility samples. However, it is clearly insufficient to describe the transport when additional processes are present, including the anomalous Hall effect (AHE) and quantum processes related to time-reversal symmetry (TRS). For the latter process, the mobility of the corresponding carrier will change as the TRS is broken by the magnetic field. Nevertheless, we discuss a comprehensive multiband model by combining the conventional multiband model and the contribution from AHE,

$$R_H = -\frac{1}{e} \frac{\sum_j \frac{n_j \mu_j^2}{1+\mu_j^2 B^2}}{\left(\sum_j \frac{n_j \mu_j}{1+\mu_j^2 B^2}\right)^2 + \left(\sum_j \frac{n_j \mu_j^2}{1+\mu_j^2 B^2}\right)^2 B^2} + \frac{R_0 \tanh \frac{B}{B_{AHE}}}{B} \tag{S16}$$

where the first term describes the conventional multiband effect with $j = 1,2,3 \ldots$ labeling subband index, and the second term describes the AHE effect as in reference [S8]. Below, we focus on the $R_H$ data shown in the main text, especially the dip-to-peak transition in the low-field region and discuss three possible scenarios. We check the goodness of fitting in these scenarios by using the Levenberg-Marquardt nonlinear curve fitting algorithm with constraint $R_S^{-1}(B=0) = \sum e|n_j u_j|$.

1) Multiband model with hole-to-electron transition (2e+e/h)

The Rashba model indicates the existence of a Dirac point marking hole-to-electron transition, which can naturally explain the dip-to-peak feature in $R_H$. We test this scenario by including 3 bands with two electron bands (by restricting $n_1, n_2 > 0$) and one hole band (by allowing $n_3$ to change sign upon gating) and drop the AHE term (by setting $R_0 = 0$). However, no reliable fitting can be found in the dip-to-peak transition range of $V_{bg}$, as shown in Fig. S9(a), suggesting this scenario is not responsible for the transport data.

2) Multiband model with AHE (2e+AHE)

The dip feature in $R_H$ has been studied in detail by Gunkel et al. in reference [S8] by introducing AHE to account for the dip feature in the Bell shape background in $R_H$ in NGO/STO samples. Although our samples show no sign of magnetic hysteresis (Fig. S8), we still test this scenario by including 2 electron bands and AHE in Eq. (S16). As a result, no reasonable fitting parameters can be found either (Fig. S9(b)), suggesting AHE is unlikely to be present in our samples.

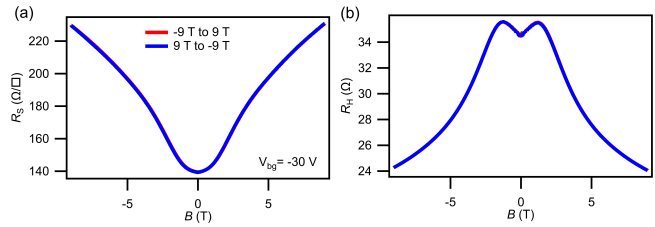

**Fig. S8.** No hysteresis is observed during consecutive magnetic field sweeps from -9 T to 9 T, then back to -9 T in both (a) $R_S$ and (b) $R_H$.

3) Multiband model with quantum corrections (2e+Q)

The quantum process that involves TRS protection, e.g., the WAL process, effectively avoids scattering and increases carrier mobility through destructive interference of electron wavefunctions. Since the magnetic field destroys TRS, this mobility enhancement is suppressed with increasing magnetic field. For the TRS-protected transport on FDWs described in the main text, it is thus plausible to assume that a magnetic field dependent mobility. The best fitting that faithfully matches the data is when an exponential dependence of mobility and magnetic field is assumed. Here we include 2 electron bands $(n_1, \mu_1)$ and $(n_2, \mu_2)$, and a quantum corrected band $(n_3, \mu_3)$ with mobility $\mu_3(B) = \mu_3^0 \exp(-\frac{B}{B_c})$. As shown in Fig. S9(c), the fitting shows excellent agreement with the data, suggesting the validity of TRS-protected transport.

To justify the fit goodness of different models, we use the root mean square error RMSE=$\sqrt{\frac{1}{n}\sum_{i=1}^{n}(y_i - \hat{y}_i)^2}$, where $y_i$ is the $R_H$ data and $\hat{y}_i$ is the predicted $R_H$ value, to qualify the fit error. The RMSE of the different model fittings are shown in Fig. S10. The fitting process of the multiband model with quantum corrections (2e+Q) yields the smallest RMSE, demonstrating that it provides the best fit to the data.

We note the quenching of the TRS-protected subband causes the overall Hall mobility to quickly drop from over 20,000 cm$^2$/Vs at low fields to ~2,000 cm$^2$/Vs at high magnetic fields. This reduced mobility can potentially explain that the QOs are not observed at the high magnetic field.

We also note the reference [S9] describes that a universal Lorentzian scaling law exists at the LAO/STO interface related to the Lifshitz transition from Ti $d_{xy}$ orbital to $d_{xz}$, $d_{yz}$ orbitals. We performed a similar analysis of the Hall coefficient data by fitting to $R_H = R_\infty + \frac{R_0 - R_\infty}{1 + (B/B_W)^2}$, where $R_0$ and $R_\infty$ is the value of $R_H$ at zero and infinite magnetic fields, respectively, and $B_W$ is the characteristic field. The fitting results (Fig. S11) show scaled $R_H$ of $V_{bg}$ do not fall under a universal scaling law, which suggests that the orbital characteristics of our samples are different from regular low-mobility LAO/STO samples.

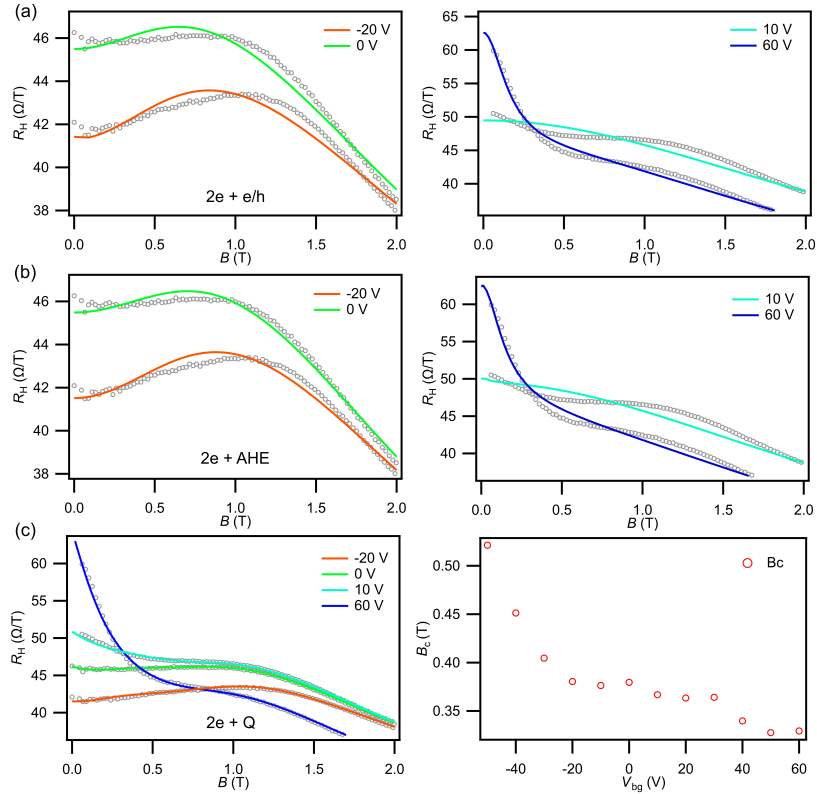

**Fig. S9.** $R_H$ **fitting by the multiband model with (a) hole-to-electron transition (2e+e/h), (b) AHE (2e+AHE) and (c) quantum corrections (2e+Q).** $R_H$ data (grey circles) and fitting results of 2e+e/h model and 2e+AHE model for -20 V (orange line), 0 V (green line), 10 V (cyan line), and 60 V (blue line) $V_{bg}$ show clear deviation in the low magnetic field region. While fitting results of 2e+Q model shows excellent agreement. The best-fitting parameters are shown in Fig. 4 in the main text. The right (c) shows the extracted characteristic magnetic field in the field dependence of $\mu_3(B) = \mu_3^0 \exp(-\frac{B}{B_c})$.

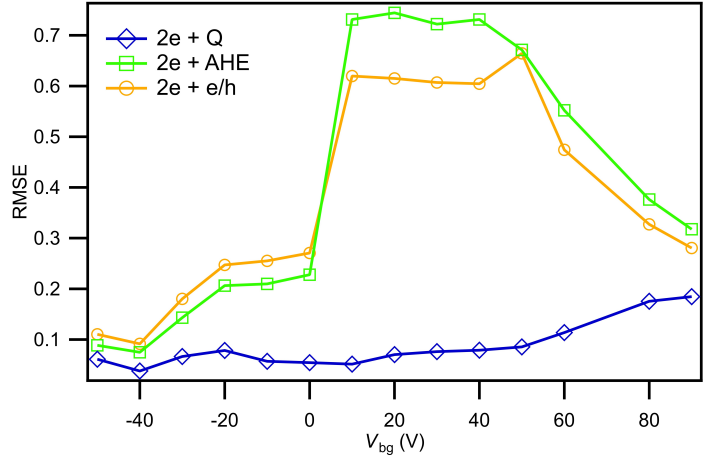

**Fig. S10. Goodness of fit.** The RMSEs of fitting the gate-tunable $R_H$ data are shown for three models: 2e+h/e (red circle), 2e+AHE (green square), and 2e+Q (blue diamond). The quantum-corrected model (2e+Q) shows the best fitting.

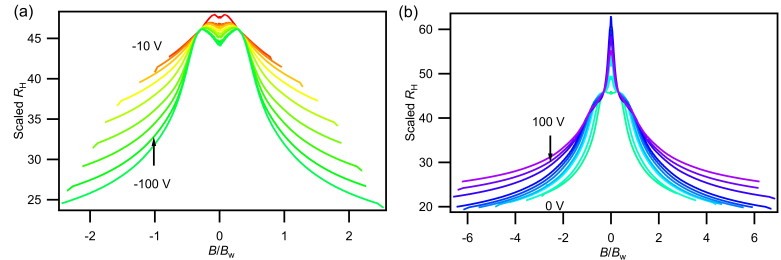

**Fig. S11. Failure of the universal Lorentzian scaling.** (a) and (b) show scaled $R_H$ of curves over $-100\,V < V_{bg} < -10\,V$ and $0\,V < V_{bg} < 100\,V$ ranges, respectively.

## CORRELATED PHASE DIAGRAM

Sample A was cooled to ~100 mK to study the relationship between superconductivity and anomalous quantum oscillations. Sample A is superconducting under relatively low $V_{bg}$ conditions. Fig. S12(a) shows the data of temperature-dependent 4 terminal resistance (in configuration A) and quantum oscillations at $T$=100 mK as a function of back gate voltages. The critical temperature $T_c$ is defined by the 10% drop of the normal resistance at 230 mK. According to the evolution of $T_c$ as a function of $V_{bg}$, superconductivity almost disappears over 20 V while quantum oscillations gradually begin to emerge, indicating a transition between superconductivity and quantum oscillations takes place over 20 V~50 V. The phase diagram of sample A as a function of temperature and $V_{bg}$ is shown in Fig. S13.

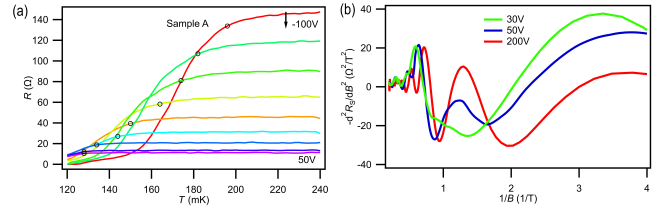

**Fig. S12. Gate-tunable (a) superconductivity and (b) quantum oscillations of samples A.** Gate voltages vary from -100 V to 50 V.

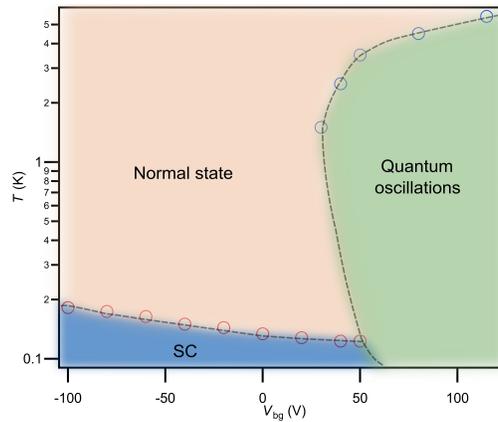

**Fig. S13. Correlated phase diagram of sample A as a function of temperature and $V_{bg}$.** The transition from superconductivity to anomalous quantum oscillations takes place at a $V_{bg}$~50 V, suggesting possible existence of a quantum critical point. The error bar (~20 V) in $V_{bg}$ marks the transition range of oscillations from emerging to fully developed.

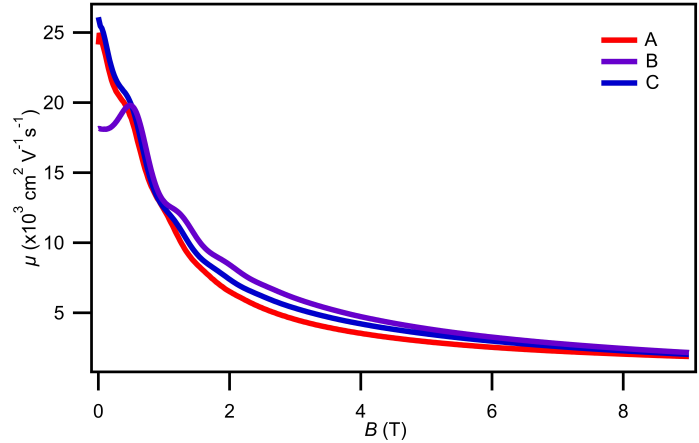

**Fig. S14. Magnetic field dependence of low-field Hall mobility (weighted average of all subbands) of sample A-C.** The mobility quickly drops from over 20,000 cm$^2$/Vs to only a few thousand with increasing magnetic field.